\documentclass{rspublic}

\usepackage[round]{natbib}
\usepackage{graphicx}
\usepackage{amssymb}

\graphicspath{{../figs_final/}}

\newcommand{\micron}{\,$\mu$m} 				
\newcommand{\dc}{\ensuremath{^{\circ}\mathrm{C}}} 	
\newcommand{\wmm}{\,W\,m$^{-2}$}			
\newcommand{\CO}{\,CO$_2$\:}

\newcommand{\gi}[1]{\textbf{\textit{#1}}}  

\title[The Runaway Greenhouse]{The Runaway Greenhouse: implications for future climate change, geoengineering and planetary atmospheres}
\author[Goldblatt \& Watson]{Colin Goldblatt$^1$ and Andrew J. Watson$^2$}
\affiliation{$^1$School of Earth and Ocean Sciences, University of Victoria, PO~Box~3065~STN~CSC, Victoria, British~Columbia, V8W 3V6, Canada (czg@uvic.ca). Previously: Astronomy Department \& Virtual Planetary Laboratory, University of Washington, Box 351580, Seattle, WA 98195, USA. $^2$School of Environmental Sciences, University of East Anglia, Norwich, NR4~7TJ, UK (a.watson@uea.ac.uk).}

\begin{document}

\maketitle

\begin{abstract}
{Runaway Greenhouse, Radiation Limit, Water Vapour, Dangerous Climate Change, Tipping Point, Venus}
The ultimate climate emergency is a ``runaway greenhouse'': a hot and water vapour rich atmosphere limits the emission of thermal radiation to space, causing runaway warming. Warming ceases only once the surface reaches $\sim$1400\,K and emits radiation in the near-infrared, where water is not a good greenhouse gas. This would evaporate the entire ocean and exterminate all planetary life. Venus experienced a runaway greenhouse in the past, and we expect that Earth will in around 2 billion years as solar luminosity increases. But could we bring on such a catastrophe prematurely, by our current climate-altering activities? Here we review what is known about the runaway greenhouse to answer this question, describing the various limits on outgoing radiation and how climate will evolve between these. The good news is that almost all lines of evidence lead us to believe that is unlikely to be possible, even in principle, to trigger full a runaway greenhouse by addition of non-condensible greenhouse gases such as carbon dioxide to the atmosphere. 
However, our understanding of the dynamics, thermodynamics, radiative transfer and cloud physics of hot and steamy atmospheres is weak.
We cannot therefore completely rule out the possibility that human actions might cause a transition, if not to full runaway, then at least to a much warmer climate state than the present one. High climate sensitivity might provide a warning. If we, or more likely our remote descendants, are threatened with a runaway greenhouse then geoengineering to reflect sunlight might be life's only hope. Injecting reflective aerosols into the stratosphere would be too short lived, and even sunshades in space might require excessive maintenance. In the distant future, modifying Earth's orbit might provide a sustainable solution. The runaway greenhouse also remains relevant in planetary sciences and astrobiology: as extrasolar planets smaller and nearer to their stars are detected, some will be in a runaway greenhouse state.
\end{abstract}

\section{Introduction}



We live in a time of rapid climate change. The warming that Earth is experiencing now due to emissions of greenhouse gases from fossil fuel burning is unprecedented in human history, and the pace of the change is very rapid compared to most past climate change as deduced from palaeoclimate records. It is natural to ask just how bad the problem could be: could humanity's collective failure to control our influence on climate lead to warming so cataclysmic as to render the planet uninhabitable? 

Wide attention has recently been drawn to such a concern by one of our most eminent climate scientists, James Hansen: 
``\ldots if we burn all reserves of oil, gas, and coal, there's a substantial chance that we will initiate the runaway greenhouse. If we also burn the tar sands and tar shale, I believe the Venus syndrome [the runaway greenhouse] is a dead certainty'' \citep[][p. 236]{hansen-09}.
In this paper, we begin by reviewing the physical basis of the runaway greenhouse in order to directly address this issue. We then consider how geoengineering could be appropriate to address a threat such as a moist or runaway climate state. Lastly, we discuss the runaway greenhouse for other planets. 

It is a common misconception that the runaway greenhouse is a simple extension of the familiar water vapour feedback. As the planet warms more water evaporates. Water vapour is a greenhouse gas, so this enhances the warming (a positive feedback). Accelerating this, by implication, would boil the entire ocean. Historically, this possibility was discussed by \citet{sagan-60} and \citet{gold-64}, but some of the earliest modern-era climate models \citep{mw-67} showed that while water vapour feedback is an important positive feedback for Earth, it is not a runaway feedback. 

In fact, the physics of the runaway greenhouse is rather different to ``ordinary'' water vapour feedback. There exist certain limits which set the maximum amount of outgoing thermal (longwave) radiation which can be emitted from a moist atmosphere. In the ordinary regime in which Earth resides at  present, an increase in surface temperature causes the planet to emit more radiative energy to space, which cools the surface and maintains energy balance. However as a limit on the emission of thermal radiation is approached, the surface and lower atmosphere may warm, but no more radiation can escape the upper atmosphere to space. This is the runaway greenhouse: surface temperature will increase rapidly, finally reaching equilibrium again only when the surface temperature reaches around 1400\,K and emits radiation in the near-infrared, where water vapour is not a good greenhouse gas. Along the way, the entire ocean evaporates. 

\citet{simpson-27} was the first to identify a radiation limit, finding a tropospheric limit, which was seen as a paradox. 
\citet{komabayashi-67} and \citet{ingersoll-69} both analytically derived stratospheric limits, which gave rise to the modern concept of the runaway greenhouse. 
\citet{rasool-debergh-70} applied a grey model to Venus.
Spectrally resolved numerical calculations of the full runaway were performed by \citet{pollack-71}, \citet{watson-ea-84}, \citet{abe-matsui-88} and \citet{k-88}. Over two decades later, these 1980s era calculations have yet to be substantially revised. 
\citet{nakajima-ea-92} used a grey radiative--convective model to develop the theoretical framework of the various radiation limits, which is followed here. 
\citet{pierrehumbert-95} examined the local runaway greenhouse in Earth's tropics, highlighting the importance of dry air columns for cooling. 
\citet{renno-97} studied the transition towards runaway conditions (up to 330\,K) with a radiative convective model with an explicit hydrological cycle, finding multiple equilibria. 
Pujol \citep{pujol-north-02,pujol-fort-02,pujol-north-03} and colleagues elaborated a semi-grey radiative-convective model finding, for example, the existence of multiple equilibria during warming.
\citet{sugiyama-stone-emanuel-05} examined the role of relative humidity.
\citet{ishiwatari-ea-02} conducted the only study of the runaway greenhouse in a General Circulation Model (GCM) to date, using a simplified model with the grey radiative transfer scheme from \citet{nakajima-ea-92}. \citet{ishiwatari-ea-07} compared these GCM results with an energy balance model to examine climate states from low-latitude glaciation to the runaway. 

\section{Physics of the Runaway Greenhouse}\label{s-phys}

In this section we review the physics of the runaway greenhouse in detail. Our theoretical framework follows \citet{nakajima-ea-92}. Whilst \citet{nakajima-ea-92} was very formal and mathematical, our description here is more qualitative.
We begin by describing three different ``limits'' on the outgoing longwave radiation. Whilst we find it useful to group these together under a single umbrella term, we emphasize that the physical character of them is rather different. Limit 1, the \textit{black body upper limit} is the maximum radiation that any planet of given surface temperature can emit; this varies with temperature. It does not lead to a runaway greenhouse, but is included for context. Limit 2, the \textit{moist stratosphere upper limit} is the maximum amount of radiation which can be transferred by a moist stratosphere; this is invariant with temperature, so can lead to a runaway greenhouse. Limit 3, the \textit{moist troposphere asymptotic limit} is the amount of radiation which a thick, pure water vapour atmosphere will emit; the radiation from any thick water-rich atmosphere will tend towards this asymptotic limit. It is invariant with temperature, so can lead to a runaway greenhouse. We then describe the changes which occur when increasing temperature in both runaway and moist greenhouses. 

Understanding the runaway greenhouse requires familiarity with many aspects of atmospheric radiative transfer and thermodynamics. Given the interdisciplinary interest in this topic, we summarise the necessary background material in \ref{ss-techbg}. 
For those who skip \ref{ss-techbg}: most of our discussion refers to an atmosphere transparent to solar (shortwave) radiation and grey with respect to thermal (longwave) radiation. Longwave optical depth is measured from the top of the atmosphere. Most emission to space is from where optical depth is of order unity, so emission depends on the temperature structure in this region. In the context of Earth, we refer to water vapour as a condensible greenhouse gas, carbon dioxide as a non-condensible greenhouse gas and radiatively inactive gases as background gas. 

\subsection{Radiation Limit 1: optically thin atmosphere / black body upper limit}



\begin{figure}
 \includegraphics[width=\textwidth]{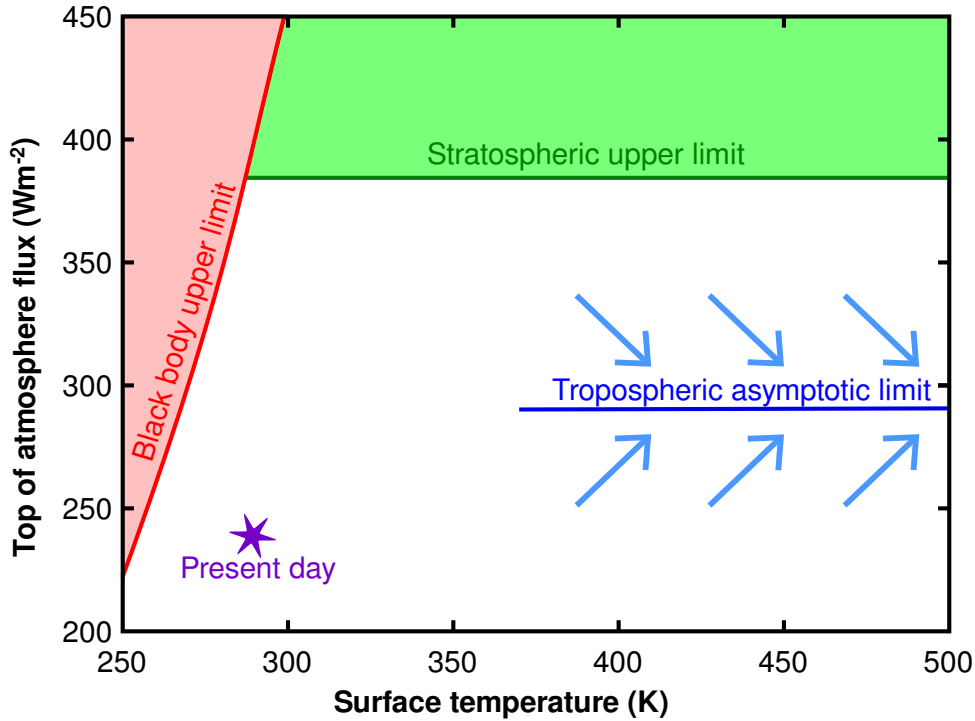}
\caption{Radiation limits (solid lines) as a function of surface temperature, after \citet{nakajima-ea-92}. Inaccessible regions are shaded. All of the white area can, in theory be occupied depending on the amount of non-condensible greenhouse gas (e.g. carbon dioxide) present, but at higher temperatures outgoing flux will tend towards the tropospheric limit, as indicated by arrows. Present day mean surface temperature is 289\,K with an outgoing thermal flux of 239\wmm. Note that the change in temperature with flux can be seen as the climate sensitivity, so the horizontal lines certain radiation limits may alternatively be seen as infinite climate sensitivity and cause a runaway greenhouse.} \label{f-runawayschematic_labels}
\end{figure}
For an airless body (the Moon for example), or an atmosphere with no greenhouse gases or clouds, radiative emission from the surface will escape to space unattenuated. The flux to space at the top of the atmosphere is the black body emission from the surface. This flux, 
\begin{equation}
 F^\mathrm{\uparrow}_\mathrm{TOA (ThinLimit)} = \sigma T^4_\mathrm{surf}
\end{equation}
is a radiation limit: Adding any greenhouse gas (i.e. making the optical depth non-negligible) will cause less radiation to be emitted to space. 

In Figure \ref{f-runawayschematic_labels} we show this and other radiative flux limits as a function of surface temperature. Since the outgoing thermal flux (plotted) must balance incoming solar flux, and if we assume that the atmosphere is transparent to shortwave, we can substitute absorbed solar flux for any given planet (about 239\wmm{} for Earth) for thermal flux and thus read surface temperature.

\subsection{Radiation Limit 2: moist stratosphere upper limit (Komabayashi--Ingersoll limit)}




\begin{figure}
\includegraphics{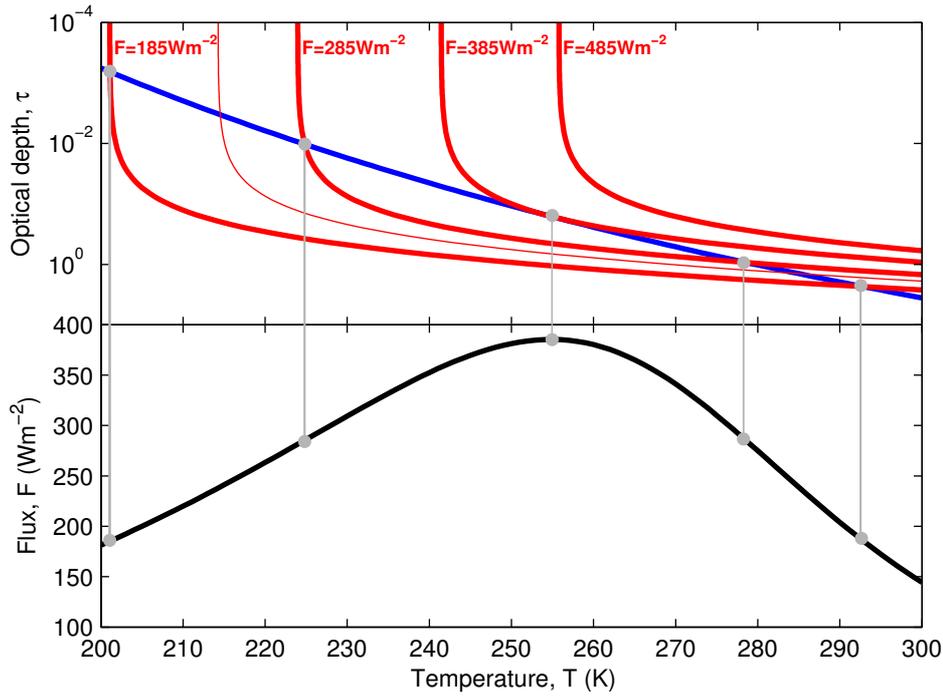} 
\caption{Top of atmosphere flux with tropopause temperature. Top panel has a reverse scale of optical depth on y-axis, so upward on plot is increasing altitude. Bold red lines are solutions of the grey atmosphere model for various fluxes, showing the temperature profile of the stratosphere. Thin red line shows current top of atmosphere flux for comparison. Blue line is optical depth of the stratosphere derived from saturated vapour pressure at the tropopause. Valid solutions for the tropopause are at intersections of red and blue lines. Bottom panel shows resulting change in top of atmosphere flux with temperature. Optical depth and grey atmosphere equations are
$\tau_\mathrm{trop} = k_v e_s(T_\mathrm{trop}) \frac{1}{g} \frac{M_v}{\bar{M}} $ 
and
$\sigma T^4 = \frac{1}{2} F^\uparrow_\mathrm{TOA} \left(\frac{3}{2} \tau + 1 \right)$
where $k_v$ is a grey absorption cross-section for water vapour, $e_s(T_\mathrm{trop})$ is the saturation vapour pressure at the tropopause temperature, $g$ is acceleration due to gravity, ${M_v}/{\bar{M}}$ is the ratio of the molar masses of water vapour and the mean air and $T$ and $\tau$ are any temperature and optical depth pair on the profile \citep{ingersoll-69,nakajima-ea-92}.} \label{f-KImatch}
\end{figure}
Historically, this is the limit discovered by the first workers to explicitly describe the runaway greenhouse \citep{komabayashi-67, ingersoll-69}, though as we will describe later, this limit is effectively impossible to achieve in practice on Earth. This limit is also the least intuitive.

Consider an atmosphere with a convective troposphere and a radiative equilibrium stratosphere, the boundary between which is the tropopause. With the atmosphere in contact with an ocean, the troposphere will be moist; to calculate a radiation limit we assume that it is saturated with water vapour. Assuming a constant mixing ratio of stratospheric water vapour, the mass of this can be calculated directly from the partial pressure of water at the tropopause, which is the saturation vapour pressure for the tropopause temperature. Optical depth depends, to first approximation, on the mass of absorber so the optical depth of the stratosphere is set by the amount of water vapour it contains, which is set by the tropopause temperature.

In radiative equilibrium, the upward longwave flux must equal the downward shortwave flux throughout. Two consequences of this are that the temperature increases with optical depth (this increase is rapid once optical depth $\tau > 0.1$) and that a higher absorbed shortwave flux will mean that the stratosphere is warmer throughout.

A limit on outgoing radiation from the stratosphere arises because the radiative equilibrium temperature profile needs to match the optical depth of the stratosphere, which is set by the saturation vapour pressure at the tropopause.
Intuitively, one might think first of the tropopause temperature then of the associated outgoing flux, but the radiation limit is easier to demonstrate by starting with the outward flux. For some given flux, we follow the radiative equilibrium temperature--optical depth profile until it intersects the temperature--optical depth curve derived from the saturation vapour pressure (figure \ref{f-KImatch}). The intersection gives the tropopause .
If the outgoing longwave flux is large, no solution can be found---the radiation limit is then the highest flux for which a solution can be found, around 385{\wmm} here.


One can understand this with a thought experiment in which we warm the tropopause. A cool tropopause means little water in the stratosphere, a small optical depth, and only a small flux is required to maintain radiative equilibrium. Increasing the tropopause temperature increases the amount of water vapour and hence the optical depth of the stratosphere, requiring a larger outgoing flux to maintain equilibrium. We can think of this intuitively as a stable regime: the consequence of increasing temperature is an increased outgoing flux, which would maintain equilibrium. This regime operates whilst the stratosphere is optically thin ($\tau < 0.1$), up to a tropopause temperature of about 255\,K. At $\tau \approx 0.1$, the rate of increase of radiative equilibrium temperature with respect to optical depth becomes larger than the rate of increase of the tropopause temperature with respect to optical depth (from the saturation vapour pressure). Consequently, when tropopause temperature is increased, the new equilibrium has a lower outgoing longwave flux. This should be unstable: decreasing the outgoing flux would cause further warming, further decreasing the outgoing longwave radiation. Thus we have runaway temperature increase\footnote{In addition, as we move to a optically thick stratosphere, our physical assumptions break down. The temperature decrease with height would exceed the moist adiabatic lapse rate and convection would occur, causing the tropopause to move up. A very warm tropopause is unlikely in the real world.}.

Thus, the simplest way to think of the Komabayashi-Ingersoll limit is to say that the stratosphere must be optically thin ($\tau \lesssim 0.1$), and that the radiation limit can be calculated assuming the tropopause is at $\tau \sim 0.1$.


\subsection{Radiation Limit 3: moist troposphere asymptotic limit (Simpson-Nakajima limit)}



\begin{figure}
\includegraphics{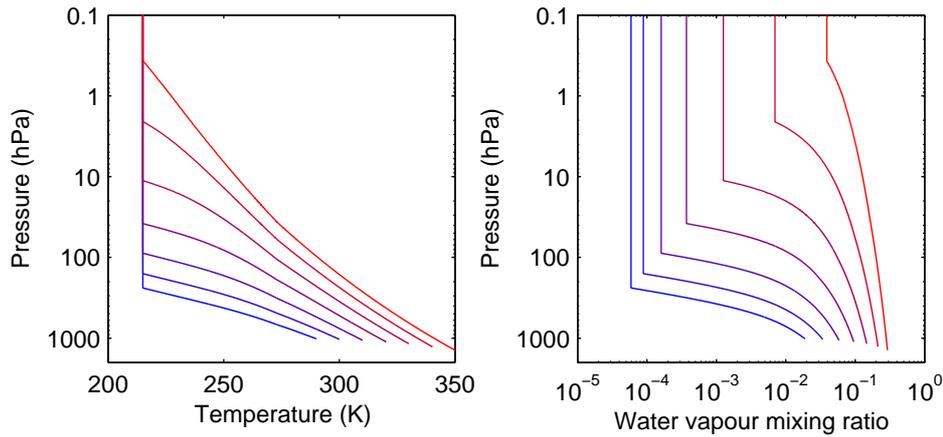}
\caption{Change in temperature structure (left panel) and moisture structure (right panel) for warming atmospheres. A background pressure of $10^5$\,Pa is assumed, equal to that of Earth's atmosphere.  }\label{f-mixratio-temp}
\end{figure}
\begin{figure}
\begin{center}
 \includegraphics{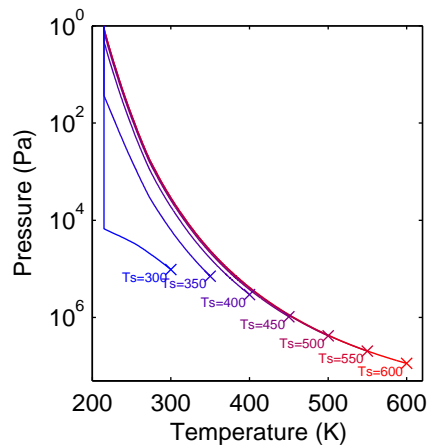}
\end{center}
 \caption{Temperature structure of the atmosphere with increasing surface temperature. A background pressure of $10^5$\,Pa is assumed, equal to that of Earth's atmosphere.}\label{f-pres-temp-tau}
\end{figure}
As the surface temperature rises, the amount of water vapour in the atmosphere increases, from water being a minor constituent (at present), to a major constituent and ultimately to the dominant atmospheric constituent (figure \ref{f-mixratio-temp}). For example, the boiling point of water at the Earth's surface pressure of $10^5$\,Pa is 373\,K (100\dc). The boiling point is, by definition, the temperature at which the saturation vapour pressure of water equals the atmospheric pressure, so with a surface temperature of 373\,K the saturation vapour pressure of water is $10^5$\,Pa, the total surface pressure would be $2 \times 10^5$\,Pa and the mixing ratio of water vapour 0.5 (now that surface pressure has increased, the boiling point would be higher than 373\,K and the entire ocean will \textit{not} boil: a liquid ocean remains until the critical point is reached at $T_c = 647$\,K). 

In general, the tropospheric temperature structure follows the moist adiabatic lapse rate. As the water vapour mixing ratio tends towards one, this simplifies to the saturation vapour pressure curve. Raising the surface temperature will evaporate more of the ocean---we can think of this as adding extra gas to the bottom of the atmosphere---but will not change the overlying atmospheric structure (figure \ref{f-pres-temp-tau}).

Effective emission to space takes place around $\tau = 1$. For the thick water vapour atmosphere we are now considering, this will be in the upper troposphere. Since the temperature structure is fixed, so too is the amount of radiative flux that can be emitted. Hence a radiation limit emerges: increasing surface temperature adds extra gas to the bottom of the atmosphere and increases surface pressure, but does not change the temperature structure of the upper troposphere from which radiation is emitted to space. The radiative flux to space is decoupled from surface temperature, with the radiation limit
\begin{equation}
 F^\mathrm{\uparrow}_\mathrm{TOA (TropLimit)} \approx \sigma T^4_{\tau=1}
\end{equation}
We refer to this as the moist troposphere limit or Simpson-Nakajima limit, for \citet{simpson-27} who first found the limit and \citet{nakajima-ea-92} who described the relevant physics.\footnote{In practice, this was the limit found in previous numerical calculations \citep{pollack-71,watson-ea-84,k-88,abe-matsui-88}, but was wrongly described there as the Komabayashi-Ingersoll limit, which should correctly refer to the stratosphere. Also, \citet{ingersoll-69} noted that a convective troposphere would not be able to deliver the required flux to meet the Komabayashi-Ingersoll limit, but did not sufficiently elucidate the physics.} 

Calculating the value of this radiation limit requires numerical evaluation of a radiative transfer code. Example values are 290\wmm{} from a grey calculation \citet{nakajima-ea-92} and 310\wmm{} from a spectrally resolved calculation \citep{k-88}.

Note that a cool atmosphere which is not pure water vapour, but contains some background non-condensible gas as Earth does, will ``overshoot'' the moist troposphere asymptotic limit (figure \ref{f-runawayschematic_paths}). This is because the presence of background gas decreases the water vapour mixing ratio, increasing the moist adiabatic lapse rate (a smaller water mixing ratio permits a bigger moist adiabatic lapse rate because there is more non-condensible gas to take up latent heat of condensation from the water), such that the temperature around optical depth of unity is higher and more radiation can be emitted. Also, with a larger lapse rate but a prescribed surface temperature, the vertically averaged atmospheric temperature is lower, so the column mass of water vapour is lower \citep{nakajima-ea-92}.

Of theoretical interest, but probably not relevant to Earth, is an analogous dry adiabatic limit. In the preceding discussion, we have considered an atmosphere where the background atmosphere ($10^5$\,Pa for Earth) is orders of magnitude smaller than the amount of water vapour available ($\sim 200 \times 10^5$\,Pa for Earth), so the saturation water vapour mixing ratio in a warm atmosphere will always be big enough to make the moist adiabatic lapse rate markedly smaller than the dry adiabatic lapse rate. By contrast, an atmosphere with a very large reservoir of non-condensible gas (2 or 3 orders of magnitude more than Earth---Venus for example) would still have small water vapour mixing ratio in a warm atmosphere, so the moist adiabatic lapse rate would be no lower than the dry adiabatic lapse rate, due to the large reservoir of non-condensible gas available to absorb the latent heat released. With the atmospheric profile tending towards the dry adiabat, the resulting radiation limit is higher than the moist adiabatic limit and is almost as high as the stratospheric limit \citep{nakajima-ea-92}.

\subsection{Increasing temperature}



\begin{figure}
 \includegraphics[width=\textwidth]{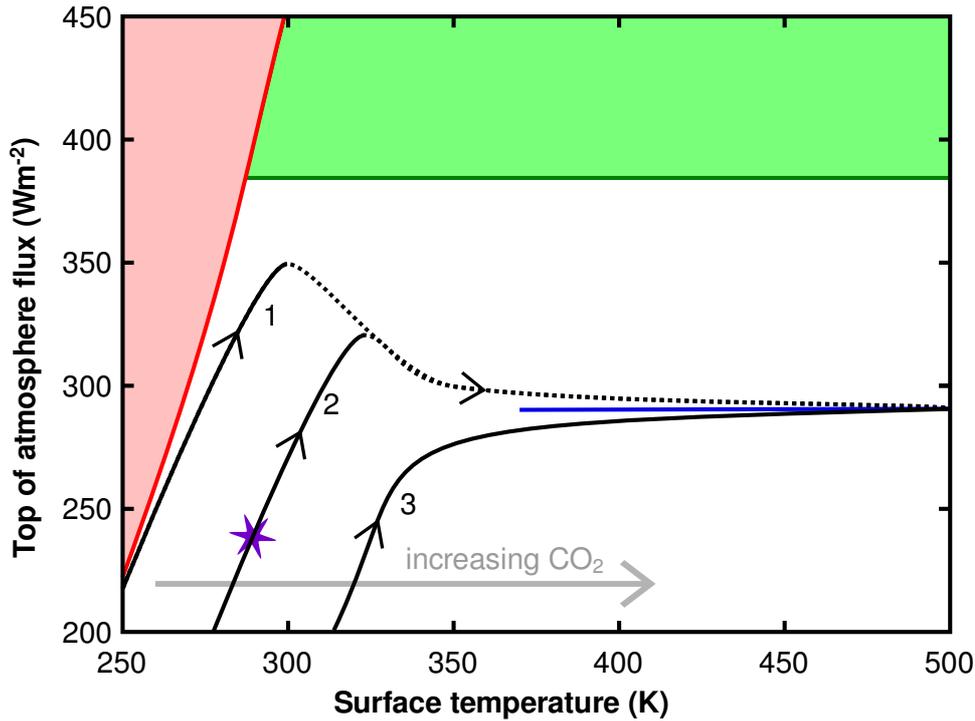}
\caption{Increase in outgoing thermal flux as a function of surface temperature, after \citet{nakajima-ea-92}. Black lines marked 1 to 3 show how the top of atmosphere flux changes with increasing surface temperature for successively higher concentrations of a non-condensible greenhouse gas (e.g. carbon dioxide). Line 2 corresponds to Earth's present amount of non-condensible greenhouse gases and lines 1 and 3 are illustrative of lower and higher concentrations, respectively. All lines are for an amount of background, non-condensible and non-radiatively active gas similar to Earth's (see figure 6 of \citet{nakajima-ea-92} for other background gas inventories).  Radiation limits are shown in colour (see figure \ref{f-runawayschematic_paths} for labels). Figure 9 of \citet{k-88}, derived from a spectrally-resolved model, has similar features. }\label{f-runawayschematic_paths}
\end{figure}
\begin{figure}
 \includegraphics[angle=270, width=\textwidth]{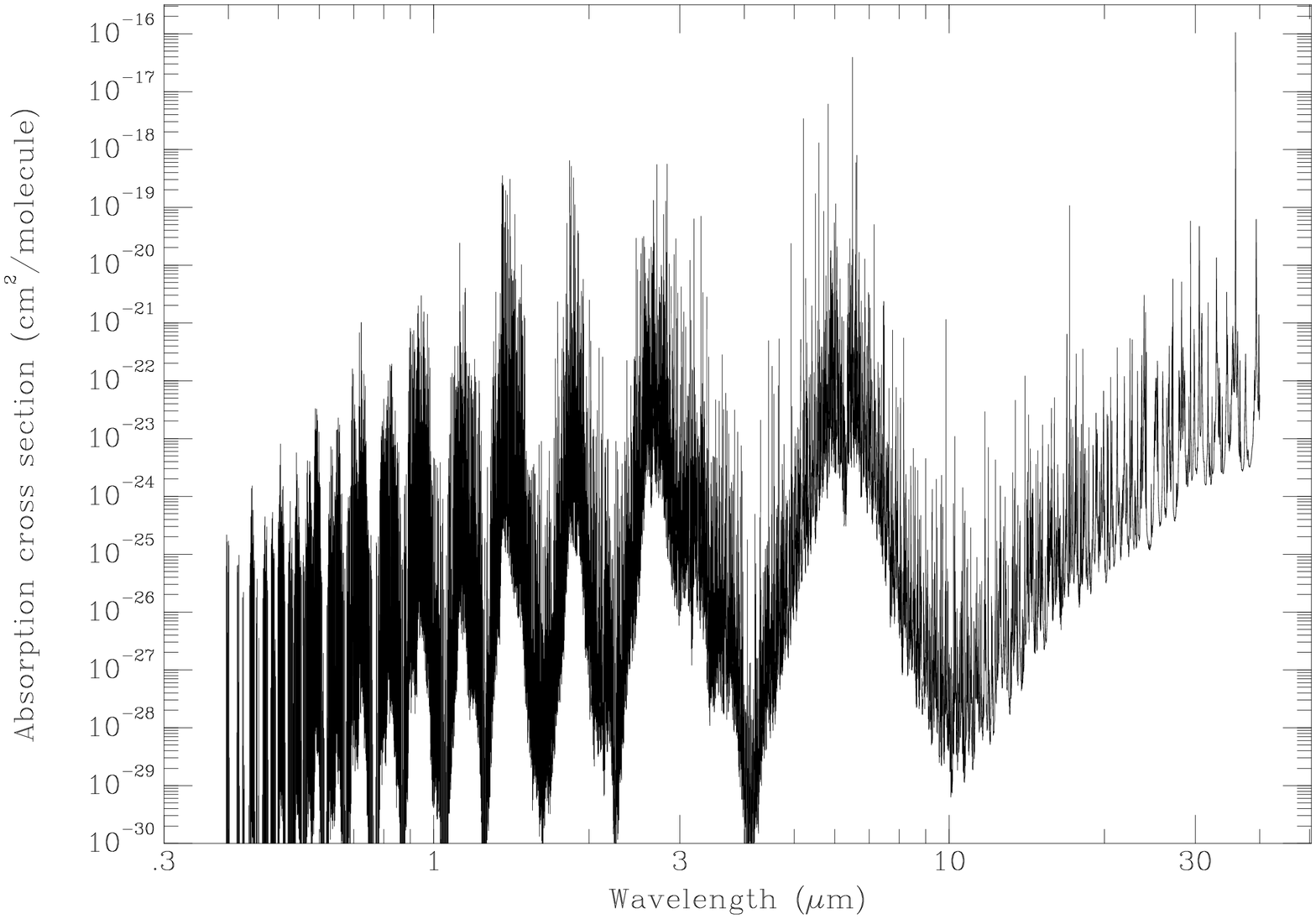}
 \caption{Absorption spectrum of water vapour 0.3 - 40 microns, shown at 220Pa and 260K. Note the `window' regions where the absorption coefficient is low and the general decline of absorption coefficient at shorter wavelength.} \label{f-wvspectrum}
\end{figure}
If we imagine increasing the solar flux absorbed by the planet, the outgoing longwave flux must increase to compensate. How will the temperature adjust to provide this increased flux? The radiation limits provide guidance, but we must interpret radiative--convective model output \citep[from][]{nakajima-ea-92} to see how these limits interact (figure \ref{f-runawayschematic_paths}). It turns out, perhaps unsurprisingly, that the path surface temperature takes depends strongly on the amount of non-condensible greenhouse gas, such as carbon dioxide, which is present.

If the atmosphere has a negligible amount of carbon dioxide, then with a low surface temperature the atmosphere is optically thin and follows the black body limit. The temperature and outgoing flux increase together in this way whilst the atmosphere is not optically thick, overshooting the moist troposphere limit. The flux approaches, but never reaches, the stratospheric limit (or dry tropospheric limit). As the atmosphere becomes warmer and richer in water vapour, the lapse rate decreases in magnitude, meaning that it is cooler around $\tau = 1$ and less flux is emitted to space. The outgoing flux then asymptotes to the moist troposphere limit (figure \ref{f-runawayschematic_paths}). Clearly, decreasing outgoing flux with increasing temperature is an unstable situation, so the onset of the runaway greenhouse is at the turning point of the flux with respect to temperature.

Now consider an increased carbon dioxide abundance. With a fixed absorbed solar flux, more \CO gives a warmer surface; this is the familiar greenhouse effect. If more \CO is present whilst we increase surface temperature the flux overshoots the moist troposphere limit less: the runaway greenhouse is initiated with a lower flux, but initiation occurs with a higher surface temperature. With a suitably large \CO inventory, there is no overshoot and the moist troposphere limit is approached monotonically (figure \ref{f-runawayschematic_paths}). In no case is the runaway initiated with a flux lower than the moist troposphere (Simpson-Nakajima) limit. 

In the simplified system, where the atmosphere is transparent to solar radiation and grey with respect to thermal radiation, the runaway greenhouse occurs when increasing surface temperature is accompanied by constant or decreasing outgoing longwave flux. This implies that the surface should warm infinitely. There is an end however, and temperature will stabilise again. To explain this, we need to relax our assumption that water vapour is a ``grey'' absorber: in reality it is quite far from being so. The absorption spectrum of water vapour (figure \ref{f-wvspectrum}) shows that there are major ``window'' regions between 8 and 12\micron{} and around 4\micron{}, where absorption is weaker than elsewhere, so radiation can escape from the surface or lower atmosphere directly to space. Today, the wavelength of maximum emission is 10\micron{} and much of the radiation that escapes Earth does so through the 8 to 12\micron{} window.  With more water vapour in the atmosphere, the 8 to 12\micron{} window will becomes optically thicker, absorbing more radiation and preventing escape to space. However, as temperature increases, the wavelength of maximum emission decreases in inverse proportion (Wien's displacement law), and radiation can escape through the shorter wavelength window regions. This allows escape from the runaway greenhouse. Numerical calculations shows that this occurs when the surface is around 1400\,K \citep{k-88,abe-matsui-88} when the corresponding wavelength of maximum emission would be 5\micron{}. Making the analogy to our original grey case, allowing emission through the 4\micron{} window can be thought of as analogous to reducing the grey absorption coefficient of the atmosphere when the surface temperature exceeds 1400\,K.

1400\,K is well above the critical temperature of water, 647\,K, at which liquid and vapour phases cease to be distinct, so the oceans all evaporate during the runaway greenhouse (the exit temperature from the runaway is basically fixed by the position of water vapour windows, so does not depend much on the volume of the oceans or whether they have all evaporated).

Relaxing the assumption that the atmosphere is transparent to shortwave radiation causes some qualitative changes as water vapour absorbs solar radiation. As the water vapour inventory increases with temperature, there is initially an increase in absorption (decrease in planetary albedo). However, as the atmosphere becomes thicker, increased Rayleigh (molecular) scattering dominates and net absorption decreases (planetary albedo increases). Consequently, the runaway greenhouse would be initiated later, as solar absorption is typically decreasing as thermal emission asymptotes \citep{k-88}. 

\subsection{The moist greenhouse}\label{s-moistgreenhouse}

So far, we have considered only the strict case of a runaway greenhouse. However, there exists an intermediate atmospheric state, the so called ``moist greenhouse'' \citep{k-88} in which water vapour is a major constituent of the troposphere and the stratosphere becomes moist (figure \ref{f-mixratio-temp}). For the troposphere, the explanation is simply the increase in saturation vapour pressure with temperature: with a 320\,K surface, for example, the water vapour mixing ratio at the surface would be 10\%. Understanding the stratosphere requires slightly more physics. We can assume that the stratospheric water vapour mixing ratio is determined by the mixing ratio at the tropopause. This is known as the cold trap; the minimum in temperature causes a minimum in satruation vapour pressure, which keeps water out of the stratosphere. However, with a warmer surface, the tropopause moves upward and to a lower pressure, but stays at a fixed temperature. Water vapour mixing ratio is the ratio of the vapour pressure to ambient pressure, so with a lower ambient pressure the tropopause mixing ratio becomes higher and the cold trap less effective. Increasing surface temperature from 290\,K to 300\,K, 310\,K or 330\,K gives increases in stratospheric water mixing ratio of factors of 1.5, 2.7 and 21 respectively.

In principle, a moist greenhouse is a stable climate---just one much hotter than today. This is not a cause for comfort however, as the differences relative to today would be profound. With a warming climate, the latitudinal temperature gradient would weaken (likely melting any year round ice) and the atmospheric circulation patterns would be quite different \citep[compare figures 8 and 9 of ][]{ishiwatari-ea-02}. Furthermore, the present stratosphere begins at around 100\,hPa and the ozone maximum is around 10\,hPa. In our simple model (figure \ref{f-mixratio-temp}) which tends to put the tropopause too low, by the time the surface temperature is 330\,K the tropopause height would have risen to engulf the ozone maximum. This could destroy the ozone layer via catalytic ozone destruction by water derived hydrogen radicals \citep{kd-80,lary-97}. Even before this, much more water in the stratosphere due to a weaker cold trap will rather change the stratospheric chemistry. A moist greenhouse would also increase climate sensitivity. More water in the stratosphere will increase the rate of hydrogen escape \citep{hunten-73} which, although not relevant on human timescales, is a major process in planetary evolution.

Getting to a moist greenhouse simply requires warming, not exceeding radiation limits: more greenhouse gases could certainly get us there. \citet{ka-86} found that somewhat over 10,000\,ppmv \CO was required. 

\section{Could anthropogenic global change cause a moist or runaway greenhouse?}

\subsection{Runaway greenhouse}

Figure \ref{f-runawayschematic_paths} illustrates how increasing the carbon dioxide inventory of the atmosphere affects the change in outgoing longwave radiation with temperature, using a grey atmosphere model. Presently, Earth's absorbed solar flux is smaller than all of the radiation limits described above and the surface temperature adjusts so that outgoing longwave radiation matches the absorbed solar flux. More carbon dioxide means that the surface must be warmer to provide the same outgoing flux---this is the familiar greenhouse effect. 

The runaway greenhouse only occurs when the outgoing longwave flux reaches a radiation limit. The fundamental point is that adding carbon dioxide does not increase the outgoing longwave flux, so cannot cause a runaway greenhouse. Whilst this result comes from simple models, a qualitatively similar result can be obtained from  spectrally resolved radiative transfer codes \citep[see figure 9 of ][]{k-88}: even 100\,bar of \CO does not give a radiation limit \citep{ka-86,k-88}.

And yet, because of the extreme seriousness of the consequences, this theoretical answer is insufficiently satisfying. Whilst we can construct simple models to describe the limits of how the atmosphere behaves (and verify them with numerical calculations to an extent), we actually have little idea of exactly how the atmosphere will change when subject to extreme heating. Is there any missed physics or weak assumptions that have been made, which if corrected could mean that the runaway is a greater risk? We cannot answer this with the confidence which would make us feel comfortable.

One major uncertainty is relative humidity. We have assumed a saturated troposphere as the end member which makes a runaway greenhouse most likely, with the aim that uncertainty here would be how much \textit{less} likely the runaway greenhouse would be. If Earth's tropics were saturated, they would be in a local runaway greenhouse, whereas in reality columns of unsaturated air allow radiation to escape \citep{pierrehumbert-95}. How relative humidity will change with warming is very poorly understood, but will have a first order effect on temperature change. Explicit inclusion of a hydrological cycle in models has given rise to multiple equilibria \citep{renno-97}, which adds an additional threat. 

Likewise, we have not discussed clouds (the physics is hard enough without them). Presently, the greenhouse effect of clouds is about half the albedo effect \citep[e.g.][]{zrlom-04}. A typical argument, following \citet{k-88}, is that in an optically thick atmosphere, the cloud greenhouse effect would diminish as non-cloud optical depth would be so high and the albedo effect would dominate further, making the runaway greenhouse less likely. This is not a watertight argument: for example, with much more water vapour aloft high clouds might become thicker and more widespread, causing net warming. We do not know what will happen. 

There are uncertainties in the radiative transfer too. The last full, spectrally resolved, treatments of the problem \citep{abe-matsui-88,k-88} are over two decades old. Much of the uncertainty relates to the so-called continuum absorption of water vapour (in window regions, both 8 to 12\micron{} and 4\micron{}, there is more absorption than one would expect considering nearby water vapour lines). This is probably caused by far-wing line absorption by water lines, which depends on self-collisions of water molecules. Hence this depends on vapour pressure squared, and is highly relevent to our problem. Understanding and empirical constraints on this continuum have improved, but remain unsatisfactory.

Are there alternate ways that the radiation limit could be reached? Could decreased albedo increase the absorbed solar radiation sufficiently? Taking the lowest radiation limit to be 285\wmm, one would need to reduce the planetary albedo from 0.3 to 0.16 to bring the absorbed solar radiation up to the radiation limit. In radiative forcing terms, this is 45\wmm. Changing all land surface to ocean (which is darker) would give a forcing of 5\wmm \citep{gz-11}. Clouds are more important: removing all low cloud would give a forcing of 25\wmm. Marine stratus depends strongly on boundary layer processes which is are not represented well in all models. Whilst albedo changes alone would unlikely be sufficient to cause a runaway greenhouse, a major reduction in cloud reflection would pose a hazard. In a warmer climate, more atmospheric water vapour would increase the amount of incoming solar radiation absorbed in the atmosphere, also decreasing planetary albedo. 

As a second alternative, we consider whether anything might lower the radiation limit. Given that the runaway greenhouse threshold depends on absorption in the 8 to 12\micron{} \citep{k-88} and 4\micron{} \citep{pierrehumbert-10} windows, it seems plausible that a greenhouse gas which absorbed strongly here (carbon dioxide does not) could lower the radiation limit. This has not been investigated. Alternatively, more extensive high cloud cover could block radiation through the windows (clouds \textit{do} approximate well to a grey absorber). If such clouds had larger particle size than present, their albedo could remain low. 

Although it seems quite unlikely, a case of more widespread and larger droplet size high clouds or good gaseous absorbers in the water vapour window and a reduction in low clouds  could, in principle, increase the absorbed solar radiation and decrease the radiation limit. Without better understanding whether such changes are possible, we cannot totally rule out a runaway greenhouse.

\subsection{Moist greenhouse}

The above considerations all apply to the case of a strict runaway greenhouse. A transition to a moist greenhouse (\S\ref{s-phys}\ref{s-moistgreenhouse}) or other hot climate state is not excluded by theory and must be seen as a potential threat until proved otherwise. In our understanding, this is the physically correct interpretation of the severe hazard of which \citet{hansen-09} warns. 

The question here is simply how much could human action increase the strength of the greenhouse effect? \citet{ka-86} found that, with carbon dioxide as the only non-condensible greenhouse gas, over 10,000\,ppmv would be needed to induce a moist greenhouse. This is likely higher than could be achieved than by burning all the ``conventional'' fossil fuel reserves---though the actual amount of fossil fuel available is poorly constrained, especially when one includes ``exotic'' sources such as tar sands (which are already being exploited). Greenhouse gases other than carbon dioxide, cloud or albedo changes could all contribute further warming. Likewise, the exhibition of multiple equilibria in the relevant temperature range \citep{renno-97,pujol-north-02} complicates matters.

\subsection{Could we predict this?}

If we did approach a moist or runaway greenhouse, how would we know? If the approach to a radiation limit is monotonic, then increased climate sensitivity might serve as a warning: close to a limit, any forcing (e.g. from more \CO) would cause a much enhanced temperature change. To put bounds on this, models used for the IPCC reports generally give a climate sensitivity to doubling \CO of around 3\,K with a likely range of 1.5 to 4.5\,K. Ensemble experiments in which uncertain parameters are varied over reasonable ranges yield a distribution of sensitivities with a long tail out to over 10\,K \citep{stainforth-ea-05}. Such high sensitivity is inconsistent with our knowledge of palaeoclimate and the model cases which provide the extremes do not seem likely (due to poor representation of contemporary climate): thus if such a sensitivity was observed, it would indicate moving to a markedly different climate state. However, figure \ref{f-runawayschematic_paths} \citep[following][]{nakajima-ea-92} and various other studies of the approach to the runaway greenhouse \citep{renno-97,ishiwatari-ea-02,pujol-north-02,sugiyama-stone-emanuel-05} indicate that there are strong non-linearities and some bifurcation points (transitions between different steady states) and multiple equilibria on this approach. The approach to a runaway greenhouse is not sufficiently understood, so it is not obvious what we should look for as warning. 

Ideally, we would want numerical climate models to robustly resolve the transition to a much hotter atmosphere. However, most such models have been developed for fairly small perturbations from the existing climate and their wider applicability may be limited by the obvious unavailability of data to tune the model to, and by simplifications made to reduce computational cost (for example, the performance of the radiative transfer schemes used in some contemporary general circulation models deteriorates at or soon after doubling carbon dioxide from present day concentrations \citep{collins-ea-06, glw-09}). Considering an atmosphere that substantial fractions of which can condense introduces many additional challenges. A new generation of model may well be needed.

\section{Geoengineering solutions to a Runaway Greenhouse}


Suppose that the runaway greenhouse did prove to be a risk, either in the foreseeable future for humanity (if our analysis above is erroneous) or for our remote descendants as solar output increases. Geoengineering would be necessary to keep Earth habitable, but how would this best be done? Three principles might guide action: (1) The threat of a runaway greenhouse would come from external heating (probably solar), not from greenhouse gases. (2) This would be a permanent threat. (3) Climate sensitivity would be very high. 

From the first point, it is clear that the geoengineering strategy would be to reduce the amount of sunlight absorbed. From the latter two points, it is equally clear that this strategy would need to be long lived and robust. Temporary reflectance-altering strategies such as addition of aerosols to the stratosphere might provide a quick fix in dire emergency, but would not answer the need for permanence. A more robust approach would be to deploy sunshades at the inner Sun-Earth Lagrange point (L1). A scheme described by \citet{angel-06} would use millions of small sunshades at the L1 point, which would block 1.8\% of the incident solar radiation (equivalent to about 6{\wmm} when averaged over the Earth). However, the small ``flyers'' would also have a limited lifetime, probably only measured in decades \citep{angel-06}, so constant maintainance would be required.

If the risk of a runaway greenhouse is distant in time, we can consider solutions beyond our present engineering capability. To counter the brightening Sun, one could move the Earth's orbit outward (astronomical engineering). Physically, this could be done by altering the orbit of an asteroid to perform gravity assists around Earth and Jupiter, transferring orbital energy from Jupiter to Earth \citep{koryckansky-ea-01}. This would be ideal as it would directly address the cause of the problem and would not require continual maintenance.

\section{The runaway greenhouse on other planets and exoplanets}

The theory relating to the runaway greenhouse was developed in the planetary sciences, concerning the evolution of Venus and planetary habitability. Reaching a radiation limit is seen as setting a hard limit on the inner edge of the circumstellar ``habitable zone'' (the region in which an terrestrial planet would likely to be habitable for life as we know it \citep{abe-93,kwr-93}) and is thus relevant for the design of exoplanet characterisation missions.  

A runaway greenhouse is caused by an increase in external heating. One heating source is solar: stellar luminosity increases with age (e.g. the sun is 30\% more luminous now than when it entered the main sequence), so time is sufficient. Alternately, if the orbit of a planet evolves inward, a radiation limit could equivilently be reached. Theoretically, a decrease in albedo could mimic increased solar flux (e.g. deglaciation from total ice cover whilst solar flux is high \citep{ishiwatari-ea-07}). The second commonly discussed heating sorce is a large astroid impact \citep[e.g.][]{abe-matsui-88}. Impact heating is effectively delivered instantly, giving an immediate transition to runaway conditions; these would be transient with the atmosphere slowly condensing afterwards. 

For insolation driven runaways (or a planet subject to relentless bolide impact), the ultimate escape from a runaway greenhouse would be loss of water to space by disassociation due to photolysis and then escape of hydrogen to space. 
Given that the atmosphere would be dominated by water vapour, hydrogen escape would be energetically limited (hydrodynamic escape) and the timescale for loss of an Earth-size ocean of water would be a few hundred million years \citep{wdw-81}.

These considerations are not entirely hypothetical: Venus' atmospheric state is thought to be explained by a runaway greenhouse and subsequent hydrogen escape. Its present atmosphere is 92 bars, consisting 96.5\% \CO and 3.5\% N$_2$. Water vapour is present at only $\sim30$\,ppmv \citep{mc-96}. Being nearer to the sun, Venus recieves nearly twice the energy from the Sun as Earth does, so will have exceeded the radiation limits at some point in its history. A hundred-fold enrichment of the deterium--hydrogen ratio on Venus relative to Earth is evidence of the subsequent hydrogen escape \citep{donahue-ea-82}.

The radiation limits vary with the acceleration due to gravity, $g$ (bigger for heavier planets). We focus here on the moist troposphere asymptotic limit (Simpson-Nakajima limit), as it is the lowest and most commonly encountered limit. To first order approximation, the optical depth of a layer depends on the mass of absorbing gas in that layer. With this approximation, the required mass of absorber in the layer between the top of the atmosphere ($\tau = 0$) and the effective emission level ($\tau = 1$) is constant. If $g$ is increased, the pressure at the bottom of this layer (at $\tau = 1$) will similarly increase ($p = \mathrm{force}/\mathrm{area} = \mathrm{mass}\cdot g/\mathrm{area}$). With the vertical temperature structure following the saturation vapour pressure curve, higher pressure at the effective emission level corresponds to a higher emission temperature, hence a higher limiting flux\footnote{Optical depth does actually depend on pressure too due to ``pressure broadening'' of absorption lines. Thus, in a more complete treatment, the limiting flux varies more slowly with $g$ than indicated here. See \citet{pierrehumbert-10} for a quantitative treatment of this.}. A Mars-size planet would have a lower limiting flux and be more susceptible to a runaway greenhouse, whereas a few Earth-mass planet would have a higher limiting flux and be less susceptible to a runaway greenhouse.\footnote{A semi-analytical treatment for the moist troposphere limit has recently been developed by \citet{pierrehumbert-10}, by substituting the vertical structure given by the saturation vapour pressure curve into a two-stream radiative transfer equation. Coefficients for radiative transfer terms are tuned to runs of a radiative transfer code. This formulation will likely be useful for ad-hoc calculations for applications of the problem (for example calculating a bound on the inner edge of the circumstellar habitable zone for variable size extra-solar planets).}

\section{Conclusion}

The runaway greenhouse is well defined in theory: As the surface warms, the atmosphere becomes optically thick with water vapour, which limits the amount of thermal radiation which can be emitted to space. If the planet absorbs more solar energy than this limit, runaway warming ensues. The practical limit on outgoing thermal radiation occurs when the atmospheric structure tends towards the saturation vapour pressure curve of water as the atmospheric composition tends towards pure water vapour, giving a limit of around 300\wmm. Earth presently absorbs around 240{\wmm} of solar radiation. Increasing carbon dioxide concentration will make surface warmer with the same outgoing thermal flux. Following this theory, we are not near the threshold of a runaway greenhouse. However, the behaviour of hot, water vapour rich, atmospheres is poorly understood and much more study of these is necessary.

In the event that our analysis is wrong, we would be left with the situation in which only geoengineering could save us. The only useful methods would be those which would reduce the amount of solar energy that the Earth absorbs. 

Lastly, we emphasise that nothing in this article should detract from the fact that anthropogenic greenhouse gas emissions are a major threat to global society and ecology. Whilst we make the technical point that these emissions are unlikely to cause a ``runaway greenhouse'', very severe and dangerous climate change is a real possibility. Abrupt change to a hot ``moist greenhouse'' state, which though not technically a ``runaway'' would nonetheless be dire, is not excluded. 
The imperative to cut greenhouse gas emissions remains. 

\begin{acknowledgements}
Thanks to Kevin Zahnle and Ray Pierrehumbert for enlightening discussions on the runaway greenhouse, to Kevin Zahnle, Naomi Vaughan, Megan Smith and Tyler Robinson for comments on the manuscript and to Jim Kasting and an anonymous reviewer for constructive reviews. Thanks to Richard Freedman for plotting figure \ref{f-wvspectrum}. CG was funded by NASA Planetary Atmospheres program (grant NNX11AC95G) and through the Virtual Planetary Laboratory Lead team of the NASA Astrobiology Institute. AJW was supported by a Royal Society Research Professorship.
\end{acknowledgements}

\appendix{Technical Background} \label{ss-techbg}

In this appendix, we summarise the background physics required to understand the runaway greenhouse, for the benefit non-experts. Key terms are emphasised for easy reference. Further background can be found in various textbooks. For example, \citet{wallace-hobbs-06} at an introductory level, \citet{cw-99} for reference on thermodynamics and \citet{pierrehumbert-10} for an accessible, intermediate level, coverage of radiative transfer and discussion of the runaway greenhouse.

We are concerned with how the atmosphere absorbs and emits radiation. Two basic properties here are the \gi{absorptivity}, $\mathcal{A}$, and \gi{emissivity}, $\varepsilon$. Absorptivity is the fraction of the radiation at any given wavelength which is absorbed by some atmospheric layer and emissivity is the fraction of the maximum radiation that the layer could emit (the black body radiation, to be discussed shortly) which is actually emitted. For any given wavelength, absorptivity equals emissivity (\gi{Kirchhoff's law}). A body which is a perfect absorber and emitter ($\mathcal{A} = \varepsilon = 1$) at all wavelengths is described as a \gi{black body}. The wavelength distribution of emitted radiation from a black body depends solely on the temperature of the body. This distribution is called the \gi{Planck function}. Two properties of this function are most relevant to the atmosphere. First, the integrated radiative emission (the radiative flux) is given $F = \sigma T^4$ (the \gi{Stefan-Boltzmann law}), where $\sigma = 5.67 \times 10^{-8}$ is the Stefan-Boltzmann constant and $T$ is the temperature. Thus the emitted radiative flux is a very strong function of temperature. Second, the peak in the Planck function (the wavelength of maximum emission) is inversely proportional to the temperature of the body (\gi{Wien's displacement law}). 

The Earth's surface and the Sun correspond roughly to black bodies with temperatures of 289\,K and 5900\,K, which have their respective Wein peaks at 10\micron{} (thermal infrared radiation) and 0.5\micron{} (visible radiation). Hence terrestrial radiation is often referred to as thermal or \gi{longwave radiation}, and solar radiation as \gi{shortwave radiation}. It is common to treat these two wavelength regions separately as their behaviour in the atmosphere is rather different. In the shortwave region, reflection dominates over absorption. Around 30\% of incident solar radiation is reflected by clouds, the surface and by molecules (\gi{Rayleigh scattering}) but only a small fraction is absorbed. Neglecting this absorption is a common approximation. In the longwave, scattering (reflection) is negligible but absorption is very important. This absorption is done both by clouds and by gases (\gi{greenhouse gases}). Whilst radiative absorption has a very rich wavelength dependence, a lot of theoretical progress can be made with the simplifying assumption that each wavelength region (long- and shortwave) can be described with a single emissivity. This is the \gi{grey atmosphere} approximation, and is used for most of the discussion in this paper.  

Now, consider the energy balance of Earth. Absorbed solar radiation must equal emitted thermal radiation at the top of the atmosphere, so we can write $\frac{S}{4}(1-\alpha) = \sigma T_e^4$, where $S = 1368$\wmm{} is the solar constant (the amount of solar energy incident at the top of the atmosphere), the factor of $1/4$ accounts for the ratio of the area the disk on which solar energy is absorbed to the Earth's surface, $\alpha = 0.3$ is the \gi{albedo} (the fraction of solar radiation reflected). We can solve this equation for the ``effective temperature'' of Earth, $T_e = 255$\,K. The surface is much warmer than this, as it is heated by the greenhouse effect.  

The general principle of the \gi{greenhouse effect} is this: the atmosphere decreases in temperature with height, so we can consider the atmosphere to be a layer which is colder than surface ($T_\mathrm{atm} < T_\mathrm{surf}$). This will absorb a fraction of the upward travelling radiation with from the surface (the fraction is the absorptivity). Recall that emissivity equals absorptivity. The atmospheric layer emits both upwards and downwards, dependant on its temperature (from the Stefan-Boltzmann law) and its emissivity, with $F^\downarrow = F^\uparrow = \varepsilon\sigma T_\mathrm{atm}^4$. Given that $T_\mathrm{atm} < T_\mathrm{surf}$, less energy is then emitted to space than would have been with no absorbing layer. Energy radiated downwards heats the surface. The strength of the greenhouse effect not only depends on the amount of greenhouse gas, but the temperature of the layer of the atmosphere from which emission occurs. The colder the emitting layer, the better the greenhouse effect.

For the discussion of the runaway greenhouse, some more technical details are necessary. A key quantity is the \gi{optical depth}, $\tau$, of a layer or the atmosphere. This depends on the number of molecules in the layer on the path the radiation takes, $N$, and some absorption coefficient for the relevant absorbing gas, $k$, thus $\tau = kN$. The amount of radiation which is transmitted through a layer depends directly on this optical depth: $F_1 = F_0 \re^{-\tau}$ (\gi{Beer's law}), where $F_0$ and $F_1$ are the incident and transmitted fluxes respectively. Hence the amount of absorption depends on the logarithm of the amount of absorbing gas. Further, most of the absorption will happen around within an order of magnitude of optical depth of one (unity): $\re^{-0.1} = 0.90$, $\re^{-1} = 0.37$, $\re^{-10} = 0.000045$. 

It is conventional to measure the optical depth of from the top of the atmosphere, with $\tau = 0$ at the top of the atmosphere, so most radiation is then emitted to space from around an optical depth of unity. The strength of the greenhouse effect thus depends on the temperature structure of the atmosphere around this level. 

We need to distinguish between \gi{condensible} and \gi{non-condensible greenhouse gasses}. Most of the greenhouse gases in Earth's atmosphere (carbon dioxide is the dominant non-condensible greenhouse gas) require a much colder temperature than is ever experienced in order to condense: we refer to these as non-condensible. Water vapour, by contrast, does condense at ambient pressures so is referred to as a condensible greenhouse gas. There is a large reservoir of the condensed phase in the oceans. The pressure of vapour which is in equilibrium (no net evaporation or condensation) is the \gi{saturation vapour pressure}. This depends exponentially on the ambient temperature. Warming Earth's surface will cause there to be much more water vapour in the atmosphere. (It is only because of the temperature of Earth's atmosphere that particular gases are condensible or non-condensible. On Mars, which is colder than Earth, carbon dioxide is condensible. On Titan, a moon of Saturn, which is colder still, water behaves more like rock does on Earth and methane is a condensible gas.). 

The structure of the atmosphere is important. We genrally consider an atmosphere with two vertical regions. The lower region, the \gi{troposphere}, is characterised by large scale vertical (convective) motions. Given the dominance of convection, the vertical temperature profile is set by the behaviour of rising air, following a \gi{moist adiabatic lapse rate} (air cools as it rises, but part of this cooling is offset by latent heat release as water condenses from the rising air). In the upper region, the \gi{stratosphere}, convection is negligible and the temperature structure is determined solely by the balance of absorbed and emitted radiation; it is said to be in \gi{radiative equilibrium}. The boundary between these layers is the \gi{tropopause}.


\begin{thebibliography}{40}
\providecommand{\natexlab}[1]{#1}
\expandafter\ifx\csname urlstyle\endcsname\relax
  \providecommand{\doi}[1]{doi:\discretionary{}{}{}#1}\else
  \providecommand{\doi}{doi:\discretionary{}{}{}\begingroup
  \urlstyle{rm}\Url}\fi

\bibitem[{Abe(1993)}]{abe-93}
Abe, Y. 1993 {Physical state of the very early Earth}.
\newblock \emph{Lithos}, \textbf{30}, 223--235.

\bibitem[{Abe \& Matsui(1988)}]{abe-matsui-88}
Abe, Y. \& Matsui, T. 1988 {Evolution of an impact generated H$_2$O--CO$_2$
  atmosphere and formation of a hot proto-ocean on Earth}.
\newblock \emph{J. Atmos. Sci.}, \textbf{45}, 3081--3101.

\bibitem[{{Angel}(2006)}]{angel-06}
{Angel}, R. 2006 {Feasibility of cooling the Earth with a cloud of small
  spacecraft near the inner Lagrange point (L1)}.
\newblock \emph{Proc. Natl. Acad. Sci. U. S. A.}, \textbf{103},
  17\,184--17\,189.
\newblock (\doi{10.1002/ep.10228})

\bibitem[{Collins \emph{et~al.}(2006)Collins, Ramaswamy, Schwarzkopf, Sun,
  Portmann, Fu, Casanova, Dufresne, Fillmore \emph{et~al.}}]{collins-ea-06}
Collins, W.~D., Ramaswamy, V., Schwarzkopf, M.~D., Sun, Y., Portmann, R.~W.,
  Fu, Q., Casanova, S. E.~B., Dufresne, J.-L., Fillmore, D.~W. \emph{et~al.}
  2006 {Radiative forcing by well-mixed greenhouse gases: Estimates from
  climate models in the Intergovernmental Panel on Climate Change (IPCC) Fourth
  Assessment Report (AR4)}.
\newblock \emph{J. Geophys. Res.}, \textbf{111}(14), D14\,317.
\newblock (\doi{10.1029/2005JD006713})

\bibitem[{Curry \& Webster(1999)}]{cw-99}
Curry, J.~A. \& Webster, P.~J. 1999 \emph{{Thermodynamics of the Atmospheres
  and Ocean}}.
\newblock London: Academic Press.
\newblock 471~pp.

\bibitem[{{Donahue} \emph{et~al.}(1982){Donahue}, {Hoffman}, {Hodges} \&
  {Watson}}]{donahue-ea-82}
{Donahue}, T.~M., {Hoffman}, J.~H., {Hodges}, R.~R. \& {Watson}, A.~J. 1982
  Venus was wet - a measurement of the ratio of deuterium to hydrogen.
\newblock \emph{Science}, \textbf{216}, 630--633.
\newblock (\doi{10.1126/science.216.4546.630})

\bibitem[{Gold(1964)}]{gold-64}
Gold, T. 1964 {Outgassing processes on the Moon and Venus}.
\newblock In \emph{The origin and evolution of atmospheres and oceans} (eds
  P.~J. Brancazio \& A.~G.~W. Cameron), pp. 249--256. Wiley.

\bibitem[{Goldblatt \emph{et~al.}(2009)Goldblatt, Lenton \& Watson}]{glw-09}
Goldblatt, C., Lenton, T.~M. \& Watson, A.~J. 2009 {An evaluation of the
  longwave radiative transfer code used in the Met Office Unified Model}.
\newblock \emph{Quart. J. Roy. Met. Soc.}, \textbf{135}(640), 619--633.
\newblock (\doi{10.1002/qj.403})

\bibitem[{Goldblatt \& Zahnle(2011)}]{gz-11}
Goldblatt, C. \& Zahnle, K.~J. 2011 Clouds and the faint young sun paradox.
\newblock \emph{Clim. Past}, \textbf{7}(1), 203--220.
\newblock (\doi{10.5194/cp-7-203-2011})

\bibitem[{Hansen(2009)}]{hansen-09}
Hansen, J. 2009 \emph{Storms of my grandchildren: The truth about the coming
  climate catastrophe and our last chance to save humanity}.
\newblock New York: Bloomsbury.

\bibitem[{Hunten(1973)}]{hunten-73}
Hunten, D.~M. 1973 {The escape of light gasses from planetary atmospheres}.
\newblock \emph{J. Atmos. Sci.}, \textbf{30}, 1481--1494.

\bibitem[{{Ingersoll}(1969)}]{ingersoll-69}
{Ingersoll}, A.~P. 1969 {The Runaway Greenhouse: A History of Water on Venus.}
\newblock \emph{J. Atmos. Sci.}, \textbf{26}, 1191--1198.

\bibitem[{Ishiwatari \emph{et~al.}(2007)Ishiwatari, Nakajima, Takehiro \&
  Hayashi}]{ishiwatari-ea-07}
Ishiwatari, M., Nakajima, K., Takehiro, S. \& Hayashi, Y.-Y. 2007 Dependence of
  climate states of gray atmosphere on solar constant: From the runaway
  greenhouse to the snowball states.
\newblock \emph{J. Geophys. Res.}, \textbf{112}, D13\,120.
\newblock (\doi{10.1029/2006JD007368})

\bibitem[{{Ishiwatari} \emph{et~al.}(2002){Ishiwatari}, {Takehiro}, {Nakajima}
  \& {Hayashi}}]{ishiwatari-ea-02}
{Ishiwatari}, M., {Takehiro}, S.-I., {Nakajima}, K. \& {Hayashi}, Y.-Y. 2002 {A
  Numerical Study on Appearance of the Runaway Greenhouse State of a
  Three-Dimensional Gray Atmosphere.}
\newblock \emph{J. Atmos. Sci.}, \textbf{59}, 3223--3238.
\newblock (\doi{10.1175/1520-0469(2002)059})

\bibitem[{Kasting(1988)}]{k-88}
Kasting, J.~F. 1988 {Runaway and moist greenhouse atmospheres and the evolution
  of \mbox{E}arth and \mbox{V}enus}.
\newblock \emph{Icarus}, \textbf{74}, 472--494.

\bibitem[{Kasting \& Ackerman(1986)}]{ka-86}
Kasting, J.~F. \& Ackerman, T.~P. 1986 {Climatic consequences of very
  high-carbon dioxide levels in the \mbox{E}arth's early atmosphere}.
\newblock \emph{Science}, \textbf{234}, 1383--1385.

\bibitem[{Kasting \& Donahue(1980)}]{kd-80}
Kasting, J.~F. \& Donahue, T.~M. 1980 {The Evolution of the Atmospheric Ozone}.
\newblock \emph{J. Geophys. Res.}, \textbf{85}, 3255--3263.

\bibitem[{Kasting \emph{et~al.}(1993)Kasting, Whitmere \& Reynolds}]{kwr-93}
Kasting, J.~F., Whitmere, D.~P. \& Reynolds, R.~T. 1993 {Habitable zones around
  main sequence stars}.
\newblock \emph{Icarus}, \textbf{101}, 108--128.

\bibitem[{Komabayashi(1967)}]{komabayashi-67}
Komabayashi, M. 1967 Discrete equilibrium temperatures of a hypothetical planet
  with the atmosphere and the hydrosphere of a onecomponent--two phase system
  under constant solar radiation.
\newblock \emph{J. Meteor. Soc. Japan}, \textbf{45}, 137--139.

\bibitem[{Korycansky \emph{et~al.}(2001)Korycansky, Laughlin \&
  Adams}]{koryckansky-ea-01}
Korycansky, D., Laughlin, G. \& Adams, F. 2001 {Astronomical engineering: A
  strategy for modifying planetary orbits}.
\newblock \emph{Astrophys. Space Sci.}, \textbf{275}(4), 349--366.

\bibitem[{Lary(1997)}]{lary-97}
Lary, D.~J. 1997 Catalytic destruction of stratospheric ozone.
\newblock \emph{J. Geophys. Res.}, \textbf{102}(D17), 21\,515--21\,526.

\bibitem[{Manabe \& Wetherald(1967)}]{mw-67}
Manabe, S. \& Wetherald, R.~D. 1967 {Thermal equilibrium of the atmosphere with
  a given distribution of relative humidity}.
\newblock \emph{J. Atmos. Sci.}, \textbf{24}(3), 241--259.

\bibitem[{Meadows \& Crisp(1996)}]{mc-96}
Meadows, V.~S. \& Crisp, D. 1996 {Ground-based near-infrared observations of
  the Venus nightside: the thermal structure and water abundance near the
  surface}.
\newblock \emph{J. Geophys. Res.}, \textbf{101}(E2), 4595--4622.

\bibitem[{Nakajima \emph{et~al.}(1992)Nakajima, Hayashi \&
  Abe}]{nakajima-ea-92}
Nakajima, S., Hayashi, Y.-Y. \& Abe, Y. 1992 {A study of the ``runaway
  greenhouse effect'' with a one-dimensional radiative--convective model}.
\newblock \emph{J. Atmos. Sci.}, \textbf{49}, 2256--2266.

\bibitem[{Pierrehumbert(1995)}]{pierrehumbert-95}
Pierrehumbert, R.~T. 1995 Thermostats, radiator fins and the local runaway
  greenhouse.
\newblock \emph{J. Atmos. Sci.}, \textbf{52}, 1784--1806.
\newblock (\doi{10.1175/1520-0469(1995)052<1784:TRFATL>2.0.CO;2})

\bibitem[{Pierrehumbert(2010)}]{pierrehumbert-10}
Pierrehumbert, R.~T. 2010 \emph{{Principles of planetary climate}}.
\newblock Cambridge, UK: Cambridge University Press.
\newblock Pp. 652.

\bibitem[{Pollack(1971)}]{pollack-71}
Pollack, J.~B. 1971 {A Nongrey Calculation of the Runaway Greenhouse:
  Implications for Venus' Past and Present}.
\newblock \emph{Icarus}, \textbf{14}, 295--306.
\newblock (\doi{10.1016/0019-1035(71)90001-7})

\bibitem[{Pujol \& Fort(2002)}]{pujol-fort-02}
Pujol, T. \& Fort, J. 2002 The effect of atmospheric absorption of sunlight on
  the runaway greenhouse point.
\newblock \emph{J. Geophys. Res.}, \textbf{107}, 4566.
\newblock (\doi{10.1029/2001JD001578,})

\bibitem[{{Pujol} \& {North}(2002)}]{pujol-north-02}
{Pujol}, T. \& {North}, G.~R. 2002 {Runaway Greenhouse Effect in a Semigray
  Radiative-Convective Model.}
\newblock \emph{J. Atmos. Sci.}, \textbf{59}, 2801--2810.
\newblock (\doi{10.1175/1520-0469(2002)059})

\bibitem[{{Pujol} \& {North}(2003)}]{pujol-north-03}
{Pujol}, T. \& {North}, G.~R. 2003 {Analytical investigation of the atmospheric
  radiation limits in semigray atmospheres in radiative equilibrium}.
\newblock \emph{Tellus A}, \textbf{55}, 328--337.
\newblock (\doi{10.1034/j.1600-0870.2003.00023.x})

\bibitem[{Rasool \& de~Bergh(1970)}]{rasool-debergh-70}
Rasool, S.~I. \& de~Bergh, C. 1970 {The runaway greenhouse and the accumulation
  of CO$_2$in the Venus atmosphere}.
\newblock \emph{Nature}, \textbf{226}, 1037--1039.
\newblock (\doi{10.1038/2261037a0})

\bibitem[{Renn\'o(1997)}]{renno-97}
Renn\'o, N.~O. 1997 {Multiple equilibria in radiative-convective atmospheres}.
\newblock \emph{Tellus A}, \textbf{49}(4), 423--438.

\bibitem[{Sagan(1960)}]{sagan-60}
Sagan, C. 1960 {The Radiation Balance of Venus}.
\newblock Tech. Rep. 32-34, Jet Propulsion Laboratory.

\bibitem[{Simpson(1927)}]{simpson-27}
Simpson, G.~C. 1927 Some studies in terrestrail radiation.
\newblock \emph{Mem. Roy. Met. Soc.}, \textbf{11}, 69--95.

\bibitem[{Stainforth \emph{et~al.}(2005)Stainforth, Aina, Christensen, Collins,
  Faull, Frame, Kettleborough, Knight, Martin \emph{et~al.}}]{stainforth-ea-05}
Stainforth, D., Aina, T., Christensen, C., Collins, M., Faull, N., Frame, D.,
  Kettleborough, J., Knight, S., Martin, A. \emph{et~al.} 2005 Uncertainty in
  predictions of the climate response to rising levels of greenhouse gases.
\newblock \emph{Nature}, \textbf{433}, 403--406.
\newblock (\doi{10.1038/nature03301})

\bibitem[{Sugiyama \emph{et~al.}(2005)Sugiyama, Stone \&
  Emanuel}]{sugiyama-stone-emanuel-05}
Sugiyama, M., Stone, P.~H. \& Emanuel, K.~A. 2005 The role of relative humidity
  in radiative–convective equilibrium.
\newblock \emph{J. Atmos. Sci.}, \textbf{62}, 2001--2011.

\bibitem[{Wallace \& Hobbs(2006)}]{wallace-hobbs-06}
Wallace, M.~W. \& Hobbs, P.~V. 2006 \emph{Atmospheric science: an introductory
  survey}.
\newblock Academic Press, 2nd edn.

\bibitem[{{Watson} \emph{et~al.}(1984){Watson}, {Donahue} \&
  {Kuhn}}]{watson-ea-84}
{Watson}, A.~J., {Donahue}, T.~M. \& {Kuhn}, W.~R. 1984 Temperatures in a
  runaway greenhouse on the evolving venus implications for water loss.
\newblock \emph{Earth Plan. Sci. Lett.}, \textbf{68}, 1--6.
\newblock (\doi{10.1016/0012-821X(84)90135-3})

\bibitem[{Watson \emph{et~al.}(1981)Watson, Donahue \& Walker}]{wdw-81}
Watson, A.~J., Donahue, T.~M. \& Walker, J. C.~G. 1981 {The dynamics of a
  rapidly escaping atmosphere: applications to the evolution of \mbox{Earth}
  and \mbox{Venus}}.
\newblock \emph{Icarus}, \textbf{48}, 150--166.

\bibitem[{Zhang \emph{et~al.}(2004)Zhang, Rossow, Lacis, Oinas \&
  Mishchenko}]{zrlom-04}
Zhang, Y.~C., Rossow, W.~B., Lacis, A.~A., Oinas, V. \& Mishchenko, M.~I. 2004
  {Calculation of radiative fluxes from the surface to top of atmosphere based
  on \mbox{ISCCP} and other global data sets: Refinements of the radiative
  transfer model and the input data}.
\newblock \emph{J. Geophys. Res.}, \textbf{109}, D19\,105.
\newblock (\doi{10.1029/2003JD004457})

\end{thebibliography}

\end{document}